\documentclass{article}

\usepackage{arxiv}
\usepackage[utf8]{inputenc} 
\usepackage[T1]{fontenc}    
\usepackage{hyperref}       
\usepackage{url}            
\usepackage{booktabs}       
\usepackage{amsfonts}       
\usepackage{nicefrac}       
\usepackage{cite}
\usepackage{booktabs}
\usepackage{multirow}
\usepackage[inline]{enumitem}
\usepackage{amsmath,amssymb,amsfonts}
\usepackage{graphicx}
\usepackage{textcomp}
\usepackage{xcolor}
\usepackage[caption=false,font=normalsize,labelfon t=sf,textfont=sf]{subfig} 
\usepackage{algorithm}
\usepackage{algpseudocode}
\usepackage{algorithmicx}
\usepackage{amsfonts,amssymb}

\usepackage{xspace}

\newcommand{\partCPUtm}{Intel\textsuperscript{\textregistered} Xeon\textsuperscript{\textregistered}\xspace}

\newcommand{\partCPU}{Intel Xeon\xspace}
\newcommand{\Intel}{Intel\textsuperscript{\textregistered}\xspace}
\newcommand{\VTunetm}{VTune\textsuperscript{\textregistered}~Amplifier\xspace}

\newcommand\blfootnote[1]{%
  \begingroup
  \renewcommand\thefootnote{}\footnote{#1}%
  \addtocounter{footnote}{-1}%
  \endgroup
}

\begin{document}
\title{Efficient Architecture-Aware Acceleration of BWA-MEM for Multicore Systems}

\author{
Vasimuddin Md, Sanchit Misra, \\Parallel Computing Lab, \\ Intel Corporation, Bangalore, India. \\ Email: \{vasimuddin.md,sanchit.misra\}@intel.com \\
\And
Heng Li, \\
Department of Biostatistics and Computational Biology, \\ Dana-Farber Cancer Institute, Boston, USA. \\
Department of Biomedical Informatics, \\ Harvard Medical School, Boston, USA. \\ Email: hli@jimmy.harvard.edu  \\
\And
Srinivas Aluru, \\
School of Computational Science and Engineering, \\ Georgia Institute of Technology, Atlanta, USA. \\ Email: aluru@cc.gatech.edu \\
}

\maketitle

\begin{abstract}

Innovations in Next-Generation Sequencing are enabling generation of DNA sequence data at ever faster rates and at very low cost. For example, the Illumina NovaSeq 6000 sequencer can generate 6 Terabases of data in less than two days, sequencing nearly 20 Billion short DNA fragments called reads at the low cost of \$1000 per human genome. Large sequencing centers typically employ hundreds of such systems. Such high-throughput and low-cost generation of data underscores the need for commensurate acceleration in downstream computational analysis of the sequencing data. A fundamental step in downstream analysis is mapping of the reads to a long reference DNA sequence, such as a reference human genome. Sequence mapping is a compute-intensive step that accounts for more than 30\% of the overall time of the GATK (Genome Analysis ToolKit) best practices workflow. BWA-MEM is one of the most widely used tools for sequence mapping and has tens of thousands of users.

In this work, we focus on accelerating BWA-MEM through an efficient architecture aware implementation, while maintaining identical output. The volume of data requires distributed computing and is usually processed on clusters or cloud deployments with multicore processors usually being the platform of choice. Since the application can be easily parallelized across multiple sockets (even across distributed memory systems) by simply distributing the reads equally, we focus on  performance improvements on a single socket multicore processor. BWA-MEM run time is dominated by three kernels, collectively responsible for more than 85\% of the overall compute time. We improved the performance of the three kernels by 1) using techniques to improve cache reuse, 2) simplifying the algorithms, 3) replacing many small memory allocations with a few large contiguous ones to improve hardware prefetching of data, 4) software prefetching of data, and 5) utilization of SIMD wherever applicable – and massive reorganization of the source code to enable these improvements. As a result, we achieved nearly $2\times$, $183\times$, and $8\times$ speedups on the three kernels, respectively, resulting in up to $3.5\times$ and $2.4\times$ speedups on end-to-end compute time over the original BWA-MEM on single thread and single socket of Intel Xeon Skylake processor. To the best of our knowledge, this is the highest reported speedup over BWA-MEM (running on a single CPU) while using a single CPU or a single CPU-single GPGPU/FPGA combination.

\noindent
Source-code: \url{https://github.com/bwa-mem2/bwa-mem2}
\end{abstract}

\copyright {2019 IEEE.  Personal use of this material is permitted.  Permission from IEEE must be obtained for all other uses, in any current or future media, including reprinting/republishing this material for advertising or promotional purposes, creating new collective works, for resale or redistribution to servers or lists, or reuse of any copyrighted component of this work in other works.}

\section{Introduction}
\label{sec:intro}


Innovations in Next-Generation Sequencing (NGS) have enabled generation of DNA sequence data at an enormous rate and very low cost. This is best exemplified by Illumina sequencers; a single Illumina NovaSeq 6000 can generate nearly 6 Terabases of data in a 44-hour run, sequencing nearly 20 billion short DNA fragments, called reads, of length $150$ base-pairs each at the cost of only \$1000 per human genome~\cite{novaseq-6000}. This has established short read sequencing using Illumina sequencers as the standard for modern genomics studies with large sequencing centers employing hundreds of such sequencers. 

The widespread use of NGS has enabled several applications -- sequencing of individual genomes, sequencing genomes of new species, sequencing RNA samples to gather digital gene expression data, sequencing metagenomic samples of communities of microbial organisms, and single cell sequencing. Each of these usually requires mapping reads to one or more reference sequences or genomes. Sequence mapping is a compute intensive step; for example, it accounts for more than $30\%$ of the overall time of the GATK (Genome Analysis ToolKit) best practices workflow~\cite{vasim-bb-bioaxiv-2018, gatk, gatk-portal}, one of the most prominent workflows for analyzing sequencing reads.


As the sequencers continue getting faster and cheaper at an exponential rate much faster than the Moore's law, we need commensurate speedups in sequence mapping. BWA-MEM~\cite{bwamem} is one of the most popular tools used for mapping short reads to reference sequences as it is able to achieve both speed and accuracy. The software already has tens of thousands of users. The volume of data necessitates the use of distributed memory systems and most of it is processed on clusters and cloud deployments with multicore processors usually being the platform of choice.

In this work, we focus on accelerating BWA-MEM on multicore systems using efficient architecture-aware implementation. Since performance on multiple sockets can be achieved by just distributing the reads equally and load imbalance is usually not an issue, our efforts are focused on single socket performance. 

An important requirement is to maintain identical output. Many genomics studies can take years to complete. For these studies, it is critical that the output of the sequence mapping tool does not change over a long period of time. Thus, we maintain identical output while accelerating BWA-MEM to allow like-for-like replacement. This makes optimizing the tool extremely challenging. The compute time of BWA-MEM is dominated by three key kernels that collectively account for over $85\%$ of the total compute time - (i) search for super maximal exact matches (SMEM) between reads and the reference sequence, (ii) suffix array lookup (SAL), and (iii) banded Smith Waterman (BSW) algorithm. There are several heuristics applied within the kernels and while connecting the kernels. The tool suffers from irregular memory accesses and irregular computation with too many branches and data dependencies making the use of SIMD extremely difficult. Attempts to make the computation or memory access more regular by modifying the algorithm usually result in change in output. 

Given the complexity of the BWA-MEM algorithm, only a few attempts have been made to accelerate it~\cite{houtgast2018, chang2016, alser2016, ahmed2015}. Most of these approaches use FPGAs or GPGPUs to accelerate one of the key kernels of BWA-MEM and report less than $2\times$ speedup over the end-to-end compute time. One of them~\cite{ahmed2015} demonstrates a $2.6\times$ performance gain on end-to-end compute time by accelerating two out of the three kernels using four FPGAs.

In this paper, we present end-to-end optimization of BWA-MEM targeting multicore systems. We demonstrate the benefits of our improvements using an \partCPUtm Skylake processor and an \partCPUtm E5 v3 processor\footnote{Intel, Xeon and Intel Xeon Phi are trademarks of Intel Corporation or its subsidiaries in the U.S. and/or other countries. Other names and brands may be claimed as the property of others. \copyright Intel Corporation}
, but our optimizations are generic and can provide performance gains on any of the modern multicore processors.

Our key contributions are as follows.
\begin{itemize}
    \item We performed rigorous architecture-aware optimizations on the three kernels by 1) designing techniques to improve cache reuse, 2) simplifying the algorithms, 3) replacing many small fragmented memory allocations with a few large contiguous ones to improve hardware prefetching of data, 4) exploiting software prefetching of data, 5) utilizing SIMD wherever applicable, and 6) applying shared memory parallelism using OpenMP. We undertook a massive reorganization of the source code to enable these improvements.
    \item We achieved $2\times$, $183\times$ and $8\times$ speedup, respectively, on SMEM, SAL and BSW kernels.
    \item Our optimizations achieve up to $3.5\times$ and $2.4\times$ speedups on end-to-end compute time over the original BWA-MEM on single thread and single socket, respectively, of \partCPU Skylake processor.
    \item Since our output is identical to the original BWA-MEM, it can seamlessly replace the original.
    \item We present detailed performance analysis of our implementation, and use the resulting insights to conduct architecture bottleneck analysis when using the Skylake processor for BWA-MEM.
\end{itemize}

To the best of our knowledge, this is the highest reported speedup over BWA-MEM while using a single CPU or a single CPU-single GPGPU/FPGA combination. 



\section{A Brief Overview of BWA-MEM}
\label{sec:background}

In what follows, we use capital letters $X$, $Y$, etc. to denote DNA sequences, modeled as strings over the alphabet $\Sigma = $ \textit{\{A,C,G,T\}} representing the four possible nucleotides (also called bases). The bases within a sequence are denoted using small letters like $a$, $b$, $c$, etc. We use $|X|$ to represent the length of sequence $X$. Let $X[i]$ denote the base at position $i$, and $X[i,j]$ ($i\leq j$) denote the substring $X[i]X[i+1]X[i+2]\cdots X[j]$. The bases `A' and `T' are considered complements of each other and `C' and `G' are complements of each other. We denote the complement of a base $c$ as $\bar{c}$.

\subsection{Short Read Sequence Mapping Problem}

Let $Q$ be a short read and $R$ be a long reference sequence. The sequence mapping problem is to find the best matches of $Q$ in $R$. For short reads, typical lengths of $Q$ range from $50$ to $250$. $|R|$ can range anywhere from a few hundred thousand (for bacterial genome) to several billions (for plant genomes). An important special case is the human genome, at $3$ billion base-pairs long.

Most popular sequence mapping tools, including BWA-MEM, use the seed and extend method.
In the seeding phase, matches for short substrings of the reads (called seeds) are found in the reference sequence while allowing for zero or very small number of differences in the match. This gives the potential locations where the entire read can match. In the extension phase, the seed matches are extended on both sides to cover the entire query, typically using a dynamic programming based algorithm, and the best matches are reported. BWA-MEM uses an FM-index of the reference sequence to perform seeding.


\subsection{FM-index}
\label{sec:background-fmindex}

\begin{figure}[!b]
    \centering
    \includegraphics[width=\columnwidth]{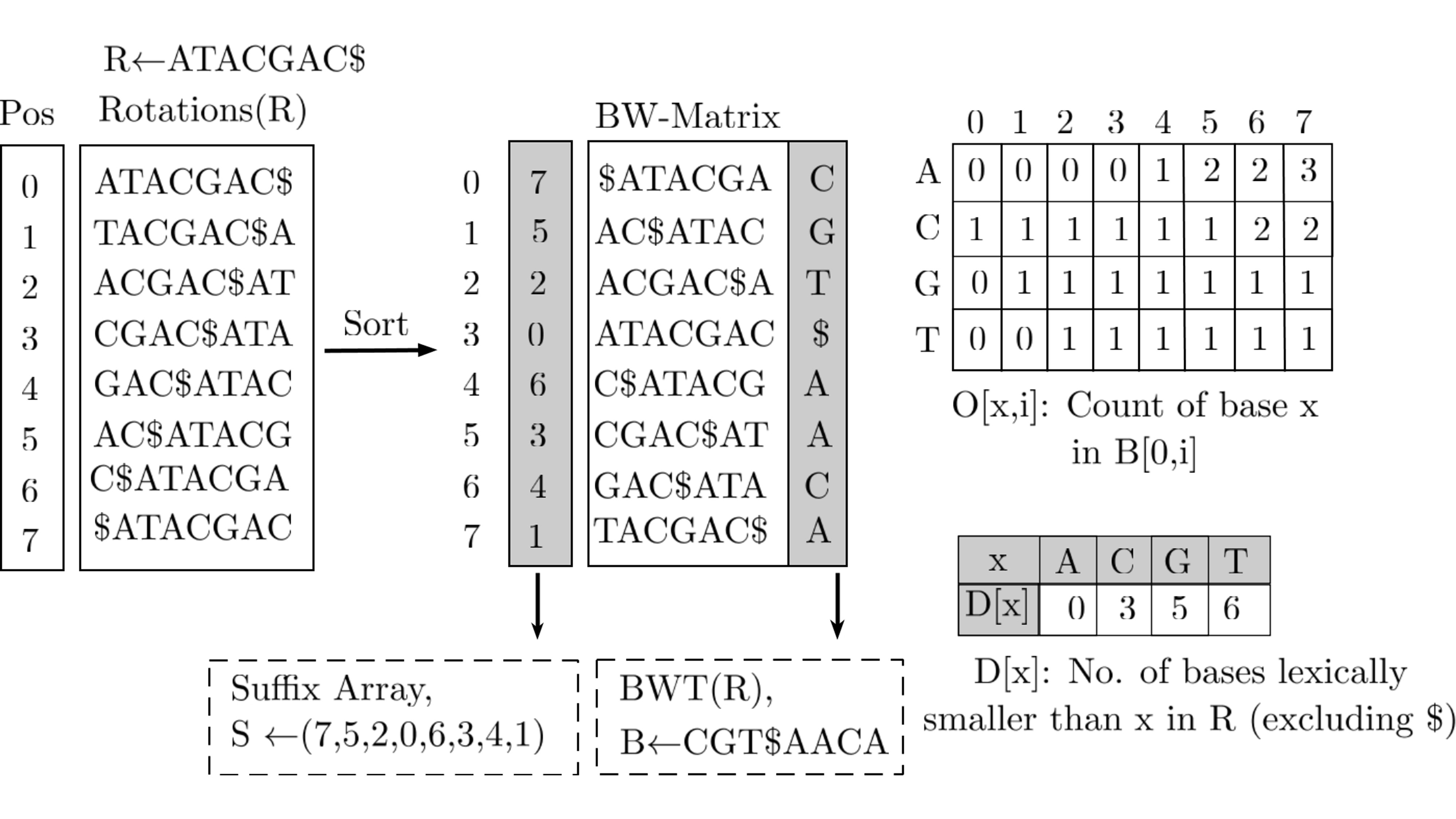}
    \caption{FM-index ($S$, $B$, $D$, $O$) and BW-matrix for reference sequence $R\leftarrow$ATACGAC\$. \$$<$A$<$C$<$G$<$T is the lexicographical ordering~\cite{misra2018}.}
    \label{fig:fm-index}
\end{figure}

The FM-index of a reference sequence is based on its Burrows-Wheeler Transform (BWT) and suffix array. Figure \ref{fig:fm-index} shows how to construct the BWT, suffix array and FM-index of an example sequence $R$. First, $R$ is appended with the character $\$ \notin \Sigma$ which is lexicographically smaller than all characters in $\Sigma$. Subsequently, all the rotations of $R$ are obtained (Rotations(R)). Lexicographically sorting the rotations gives the BW matrix. The BWT ($B$) is the last column of this matrix.
The suffix array ($S$) stores the original positions in $R$ of the first bases of these rotations, equivalent to the sorted order of the suffixes of $R$.
All the exact matches of a query are prefixes of the rotations in the BW-matrix. These matches are located in contiguous rows of the BW-matrix since the rotations are lexicographically sorted. Therefore, all the matches of a query can be represented as a range of rows of the BW-matrix. This range is called the SA interval of the query. For example, in Figure~\ref{fig:fm-index}, matches of sequence \emph{``AC''} correspond to the SA interval $[1,2]$.

To facilitate fast search of BWT, FM-index is used ~\cite{ferragina2001experimental}. It consists of $D$ and \textit{O} data structures. $D[x]$ is the number of bases in $R[0,|R|-1]$ (excluding $\$$) that are lexicographically smaller than $x \in \Sigma$, and $O[x,i]$ is the number of occurrences of $x$ in $B[0,i]$.

\subsection{BWA-MEM algorithm}
\label{sec:background-algo}

The BWA-MEM algorithm consists of the following steps.

\begin{itemize}
    \item[] {\bf SMEM:} Perform seeding by searching for super maximal exact matches (SMEM) between the read and the reference sequence. Formally, \textit{maximal exact matches} (MEMs) are exact matches between substrings of two sequences that cannot be further extended in either direction. An SMEM is a MEM that is not contained in any other MEMs on the query sequence~\cite{li2012}. BWA-MEM uses FM-index to search for SMEMs and outputs SA intervals of the SMEMs.
    \item[] {\bf SAL:} Suffix array lookup is performed to get the coordinates in the reference sequence from the SA intervals.
    \item[] {\bf CHAIN:} BWA-MEM chains the seeds that are collinear and close to each other. The chains thus formed are used to filter seeds for alignment.
    \item[] {\bf BSW:} The seeds are extended using dynamic programming (DP) based banded Smith-Waterman (BSW) alignment algorithm that computes only a diagonal band of the DP matrix. The diagonal band may shrink as the computation moves from the top row to the bottom row.
    \item[] {\bf SAM-FORM:} The algorithm concludes by formatting the alignment output in the SAM format.
\end{itemize}

\subsection{Break up of computation time spent}

\begin{table}[!b]
    \centering
    \caption{Single thread run-time profiling of BWA-MEM workflow on D1 and D4 datasets (Table~\ref{tab:app_sn_dataset}). We performed the profiling using default parameters and single-end reads.}
    \begin{tabular}{l|l|rr}
    \toprule
       Stage & Compute Blocks/Datasets & D1 & D4  \\
        \midrule
         \multirow{3}{*}{1} & SMEM &  $21.5\%$ & $44.4\%$ \\
           & SAL &  $18\%$ &  $15.5\%$\\
           & CHAIN & $6\%$ & $5.9\%$ \\
           \midrule
         \multirow{2}{*}{2} & BSW Pre-processing & $4.7\%$ & $4.9\%$ \\
           & BSW & $47.2\%$ & $26.4\%$ \\
           \midrule
         3 & SAM-FORM & $2.5\%$ & $2.9\%$\\
         \midrule
         & Total run-time & 292.73 & 182.79\\
         \bottomrule
    \end{tabular}
    \label{tab:run_profile}
\end{table}

To identify compute hotspots for acceleration,
we performed rigorous run-time profiling of BWA-MEM (version 0.7.25) by hand instrumenting the source code with trackers. Table~\ref{tab:run_profile} reports percentage of the run-time spent in various steps of BWA-MEM. 
Clearly, SMEM, SAL and BSW are the most time consuming steps accounting for $86.5\%$ and $85.7\%$ of the total run-time of BWA-MEM, on D1 and D4 datasets, respectively. Therefore, we select them as targets for acceleration.

\subsection{Potential for improvement of the key kernels}

In this section, we analyze the performance characteristics of the three kernels of BWA-MEM to identify target areas for performance improvement. To do so, we extracted the source code corresponding to each kernel from BWA-MEM and created a benchmark for each kernel. To prepare input datasets for the benchmarks, we executed BWA-MEM using read datasets and intercepted inputs to each of the kernels. The benchmarks produce the exact same output as the corresponding kernels in the original BWA-MEM.

\subsubsection{SMEM}

As shown in Section ~\ref{sec:smem-vtune}, Table ~\ref{tab:smem-counters}, the SMEM kernel executes nearly $17$ Billion instructions for $60,000$ reads; that is; nearly $285$ thousand instructions per read. The main reason for this is that BWA-MEM uses a compressed version of FM-index. Due to excessive compression, it requires a large number of instructions to compute the SA interval. Moreover, the kernel suffers from high LLC miss rate resulting in a significantly high average memory latency ($24$).

\subsubsection{SAL}

SAL kernel just retrieves the coordinate of reference sequence from the SA interval. What should have been a simple array lookup requires nearly $5000$ instructions per lookup due to the use of compressed suffix array.

\subsubsection{BSW}

Due to the irregular nature of the BSW algorithm with significant branching and short loops, it uses a scalar implementation. As we explain in Section~\ref{sec:bsw-algo}, BSW is a compute bound kernel. So, while memory access is not an issue, BSW implementation of BWA-MEM is instruction bound, due to the scalar implementation, resulting in lower performance.

\section{Modifications Applied to Entire Code}
\label{sec:opt-full-code}

\subsection{Reorganization of the Workflow}
\label{sec:opt-reorg}

\begin{figure}[!t]
    \centering
    \includegraphics[width=\linewidth]{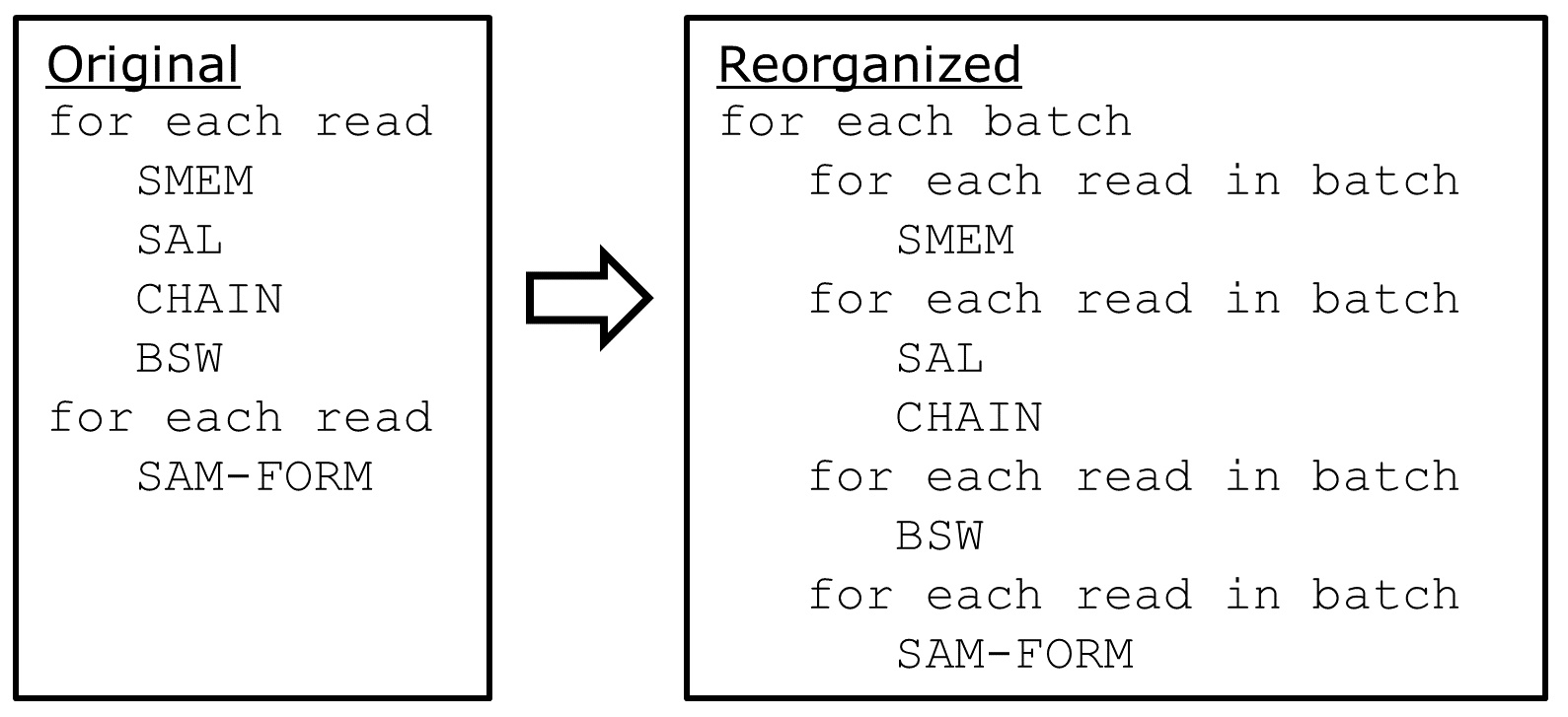}
    \caption{Reorganization of the workflow of processing of a chunk of reads.}
    \label{fig:transform}
\end{figure}

Original BWA-MEM processes a chunk of reads at a time. Figure~\ref{fig:transform} shows how we transform the workflow of processing one chunk. For each read in a chunk, the original implementation processes it through all the steps until BSW before moving on to the next read. It maintains the BSW output for the reads until the entire chunk is processed. The BSW output of each read is then converted to SAM format. Multithreading is done using \textit{pthreads} by dynamically distributing the reads across threads. On the other hand, we divide the reads in a chunk into batches. Each step is executed on all the reads in a batch before moving to the next step. Multithreading is done using OpenMP by dynamically distributing the batches across threads.

The benefit of such re-organization is that it enables the use of SIMD parallelism across different reads in a batch. We make use of this for BSW.

\subsection{Improving Inefficient Memory Allocation}

The BWA-MEM implementation allocates and de-allocates small blocks of memory frequently, which is expensive, does not provide hardware prefetcher with a long predictable pattern of memory access, and does not allow cache reuse. Instead, we allocate all the required memory once in large contiguous blocks and reuse it across the batches. Contiguous memory allocation provides benefits of hardware prefetching, reducing the memory latency. Moreover, since same buffer is used repeatedly across batches, it also improves cache reuse.

\section{Optimizations Applied to SMEM and SAL}
\label{sec:opt-smem-sal}

\subsection{FM-Index Structure}
\label{sec:opt-index}

\alglanguage{pseudocode}
\begin{algorithm}[!tb]
	\caption{$Get\_O(c,t)$: Returns $O[c,t]$}
    \label{algo:compr-occ}
	Input: Base $c$ and $O$ index $t$\\
    Output: $w$, the value of $O[c,t]$
	\begin{algorithmic}[1]
		\State $t' \gets \frac{t}{\eta}$
    	\State $y \gets t$ mod $\eta$
    	\State $w \gets O_c[t'].count(c)$
    	\State $w' \gets $ count of $c$ in first $y$ positions of $O_c[t'].bwt$
    \item[]
	\Return $w + w'$
\end{algorithmic}
\end{algorithm}

The implementation of FM-index in BWA-MEM differs
from the description in Section~\ref{sec:background-fmindex} in two ways. First, SMEM algorithm requires the FM-index of the sequence formed by concatenating the reference sequence with its reverse complement. Second, in order to reduce the memory footprint, the FM-index is compressed by dividing \textit{O} into buckets of size $\eta$. The compressed data structure $O_c$ has only $|R|/\eta$ entries. In each entry $i$, two items are stored -- 1) $count$: the count values of the bases up to bucket $i$; thus, for each base $c$, $O_c[i].count(c)$ holds the counts $O[c, i \times \eta]$, and 2) $bwt$: the substring of BWT, $B[i \times \eta,(i + 1) \times \eta - 1]$. The \textit{O} values of any position can be computed using the $O_c$ $count$ and $bwt$ (Algorithm ~\ref{algo:compr-occ}). A power-of-$2$ number is picked for the value of $\eta$ to replace the expensive division and modulo operations (lines 1 and 2) with \emph{right-shift} and \emph{bitwise AND} operations. Base $c$ and $bwt$ use 2-bit representation using $\lbrace 0,1,2,3 \rbrace$ for $\lbrace$A,C,G,T$\rbrace$. 
The count of base $c$ in the first $y$ positions of $bwt$ (line 4) is obtained by using bitwise operations.

\subsection{SMEM Search Algorithm}
\label{sec:smem-algo}

\alglanguage{pseudocode}
\begin{algorithm}[!tb]
	\caption{Backward\_Ext($(k,l,s), b$)}
    \label{algo:backward-ext}
	Input: Bi-interval $(k,l,s)$ of DNA string $X$ and a base $b$\\
    Output: Bi-interval of string $bX$
	\begin{algorithmic}[1]
        \For {$c \gets 0$ to $5$}
            \State $k_c \gets D(c) + Get\_O(c,k-1)$
            \State $s_c \gets Get\_O(c,k+s-1) - Get\_O(c,k-1)$
		\EndFor
        \State $l_0, l_1, l_2, l_3, l_4, l_5 \gets f(l, l_0, l_1, l_2, l_3, l_4, l_5, s_0, s_1, s_2, s_3, s_4, s_5)$
        \item[]
        \Return $(k_b,l_b,s_b)$
\end{algorithmic}
\end{algorithm}

\alglanguage{pseudocode}
\begin{algorithm}[!tb]
	\caption{Forward\_Ext($(k, l, s), b$)}
    \label{algo:forward-ext}
	Input: Bi-interval $(k, l, s)$ of DNA string $X$ and a base $b$\\
    Output: Bi-interval of string $Xb$
	\begin{algorithmic}[1]
	    \State $(l',k',s') \gets$ Backward\_Ext($(l, k, s), \bar{b}$)
        \item[]
        \Return $(k',l',s')$
\end{algorithmic}
\end{algorithm}

\alglanguage{pseudocode}
\begin{algorithm}[!tb]
	\caption{SMEM($X, i_0$)}
    \label{algo:smem}
	Input: String $X$ and start position $i_0$\\
    Output: Set of bi-intervals of SMEMs overlapping position $i_0$ in $X$
	\begin{algorithmic}[1]
	   \State Initialize \textit{Curr}, \textit{Prev} and \textit{Match} as empty arrays
	   \State $(k,l,s)$ $\gets$ $(D([X[i_0]], D[\overline{X[i_0]}],
	   D(X[i_0] + 1) - D[X[i_0]])$
	   \For {$i \gets i_0+1$ to $|X|$}
	       \State $\cdots$
	       \State $(k',l',s') \gets$ Forward\_Ext$((k,l,s), X[i])$
	       \If {$s' \neq s$}
	           \State Append $(k,l,s)$ to \textit{Curr}
	       \EndIf
	       \State $\cdots$
	       \State $(k,l,s) \gets (k',l',s')$
	       \State Prefetch$(O_c, l-1)$
	       \State Prefetch$(O_c, l + s-1)$
	   \EndFor
	   \State Swap \textit{Curr} and \textit{Prev}
	   \For {$i \gets i_0-1$ to $-1$}
	       \State Reset \textit{Curr} to empty
	       \For {$(k,l,s) \in$ \textit{Prev}}
	            \State $(k',l',s') \gets$ Backward\_Ext($(k,l,s), X[i]$)
	            \If {(k',l',s') is not a \textit{Match}}
	                \If {no longer matches starting at $i+1$}
	                    \State Append $(k,l,s)$ to \textit{Match}
	                \EndIf
	            \EndIf
	            \If {$(k',l',s')$ is a valid unique match}
	                \State Append (k',l',s') to \textit{Curr}
	                \State Prefetch$(O_c, k'-1)$
	                \State Prefetch$(O_c, k'+s-1)$
	            \EndIf
	       \EndFor
	       \If{\textit{Curr} is empty}
	            \State break
	       \EndIf
	       \State Swap \textit{Curr} and \textit{Prev}
	   \EndFor
	   \item[]
       \Return \textit{Match}
\end{algorithmic}
\end{algorithm}

Algorithms ~\ref{algo:backward-ext}, ~\ref{algo:forward-ext} and ~\ref{algo:smem} provide the details of SMEM search algorithm as given in ~\cite{li2012}. The dots signify details of the algorithm that are omitted as they are not relevant to this discussion. The software prefetching, introduced by us, is discussed in the next Section. Algorithm ~\ref{algo:smem} returns all the SMEMs passing through position $i_0$ in $X$. It starts at position $i_0$ and finds all the extensions in the forward direction that have matches. For all matches, it maintains the bi-interval, $(k,l,s)$, where - (i) $k$ is the starting position of the SA interval of the matching substring of $X$, (ii) $l$ is the starting position of the SA interval of the reverse complement of the matching substring, and (iii) $s$ is the size of the SA interval. For all the forward extensions, it finds the longest backward extensions that have matches to find all the SMEMs. The forward extension algorithm calls backward extension, which in turn calls $Get\_O$ that accesses $O_c$. The successive inputs $(k,l,s)$ to backward extension have no particular order or relationship. Therefore. access to $O_c$ has no spacial or temporal locality and is completely irregular. Given the huge size of $O_c$, this leads to frequent cache misses making SMEM kernel memory latency bound. 

While the compression of FM-index reduces memory footprint, there is also an additional benefit. More accesses fall into the same $O_c$ entry, improving data locality. In particular, as the matching string gets longer, the corresponding SA interval gets shorter and the likelihood of both $k$ and $k+s$ entries falling in the same cache line increases. With compressed FM-index, this should happen more frequently. The trade off is the increase in runtime for processing a base from $O(1)$ to $O(h)$.

\subsection{Applying Software Prefetching to SMEM Kernel}
\label{sec:smem-sw-prefetch}

Given the irregular memory access to $O_c$, hardware prefetching will be ineffective. However, note that each access to $O_c$ results in new values of $(k,l,s)$ that, if a match, decides the memory locations of a later access to $O_c$. We can prefetch this location using \textit{software prefetching}. Therefore, whenever new values of $(k,l,s)$ are computed that are likely to be used in the future to access $O_c$, we \textit{software prefetch} the corresponding locations of $O_c$. During forward extension, if a new match ($(k,l,s)$) is found, it is either used in the next forward extension or appended to \textit{Curr} which is used for a future backward extension. Therefore, we prefetch the corresponding locations of $O_c$. Similarly, during backward extension, any new match ($(k,l,s)$) that is added to \textit{Curr} is used for a future backward extension and, thus, we prefetch the corresponding locations of $O_c$.
However, \textit{software prefetching} will only be useful if we can hide the latency of prefetching by computation. When a match found during forward extension is used for the next forward extension, or when the number of matches in \textit{Prev} are small during backward extension, there may not be sufficient compute to hide memory latency. Since backward extension stage dominates the run time and \textit{Prev} is sufficiently large frequently enough, \textit{software prefetching} is quite useful. But it can not alleviate memory latency completely.

In order to get sufficiently large compute to hide the memory latency for every $O_c$ access, we also tried processing multiple queries simultaneously in a round robin fashion performing one call to backward extension (or forward extension) per query every time and \textit{software prefetching} according to the new match produced. However, given the intricacy of the algorithm and number of different branches possible, this approach resulted in a much larger number of instructions. The additional advantage of alleviating memory latency completely could not overcome the disadvantage of extra instructions.

\subsection{Choosing Compression Factor}

For our implementation, we choose the value of $\eta$ to be $32$ due to the following reasons.
It is better to ensure that the size of one entry of $O_c$ is less than that of one cache line. Otherwise, we will need to prefetch multiple cache lines per $O_c$ access that can cause more data to be read from memory resulting in higher memory bandwidth requirement. 

Each count needs $4$ bytes (unsigned integer). Thus, in a $O_c$ entry, $16$ bytes are consumed to store the count values. That leaves a maximum of $48$ bytes for the BWT substring. We use power-of-$2$ value for $\eta$ to prevent expensive division and modulo operations. Thus, we can only use $32$ bytes for the BWT substring. BWA-MEM uses the value of $\eta$ as $128$ and uses $2$-bits to represent each base in the BWT string, thus consuming $32$ bytes for the BWT string. However, extracting the count of a particular base from a BWT string of length $128$ requires a large number of instructions. Moreover, 2-bit representation requires significant bit manipulation making it expensive. Therefore, we opt for a one byte representation of bases to enable partial vectorization of the  computation of occurrences, thus choosing the value of $\eta$ as $32$. We perform a byte level compare using AVX2 to get a 32-bit mask containing $1$ for match and $0$ for mismatch. Consequently, we use a 32-bit popcnt instruction on the mask to get the count. We store each entry of $O_c$ in a cache aligned location for effective prefetching by adding padding of $16$ bytes to use a full cache line per entry.

\subsection{Improvements to Suffix Array Lookup}
\label{sec:opt-sal}

Given a position $i$ in the suffix array, $S$, SAL returns the corresponding coordinate in reference sequence, $j$, as shown in Equation~\ref{eq:sal}.

\begin{equation}
\label{eq:sal}
j = S[i]
\end{equation}

Similar to FM-index, BWA-MEM stores a compressed version of $S$ with compression factor of $128$. It obtains the value of $S[i]$ through an intricate algorithm requiring access to multiple locations in the compressed $S$ and FM-index. As a result, it requires nearly $5000$ instructions on an average for each $S[i]$ and a large number of memory accesses.

There is sufficient memory available to store the suffix array for most reference sequence sizes. For example, for entire human genome, we need about $48$ GB of memory. Therefore, we use the uncompressed SA and use the expression in Equation~\ref{eq:sal} to get the coordinate of reference sequence.

\section{Optimizations Applied to BSW}
\label{sec:opt-bsw}



\subsection{Banded Smith Waterman Algorithm Used in BWA-MEM}
\label{sec:bsw-algo}

The BSW algorithm used in BWA-MEM is derived from the Smith Waterman (SW) algorithm, which is a dynamic programming (DP) based sequence alignment algorithm. Given a pair of DNA sequences $X$ and $Y$, both SW and BSW compute the cells of a DP matrix $H$ of size $|X| \times |Y|$ in a row-wise manner. Here, the value $H[i,j]$ at each cell represents the alignment score between sub-sequences $X[0,i-1]$ and $Y[0,j-1]$. The key difference in the computation pattern is that, while SW computes all the cells of the DP matrix, 
\begin{enumerate*}[label=(\alph*)]
    \item BSW restricts the cell computations to within a certain band size ($\lambda$) around the main diagonal, 
    \item BSW aborts the matrix computations, if all the scores in a row are zero or the best score of the current row drops by a certain threshold from best score so far, and
    \item after computing a row, BSW adjusts the value of $\lambda$ based on number of cells with zero value from both ends.
\end{enumerate*}
The top matrix of Figure~\ref{fig:bsw} illustrates the full matrix and a diagonal band.

In BSW, computation of a cell $(i,j)$ of $H$ uses the following recursion.
\vspace{-1pt}
\begin{equation}
\begin{split}
&H[i,j] = {\max} \lbrace H[i-1,j-1] + S(X[i],Y[j]), E[i,j], F[i,j] \rbrace \\
&E[i+1,j] = {\max} \lbrace H[i,j] - \mathsf{g}_o, E[i,j] - \mathsf{g}_e, 0 \rbrace  \\ 
&F[i,j+1] = {\max} \lbrace H[i,j] - \mathsf{g}_o, F[i,j] - \mathsf{g}_e, 0 \rbrace  
\end{split}
\end{equation}

\noindent Where, $\mathsf{g}_o$ and $\mathsf{g}_e$ are gap open and extension penalties.
$S(a,b) = m$ (match score), if $a=b$, else $S(a,b) = m'$ (mismatch score). For $1 \le i \le |X|$, $1 \le j \le |Y|$, the matrices are initialized as,

\vspace{-1pt}
\begin{equation}
    \begin{split}
    H[i,0] &=  {\max }\lbrace 0, H[0,0] - (\mathsf{g}_o + \mathsf{g}_e . i) \rbrace, \\
    H[0,j] &= {\max }\lbrace 0, H[0,0] - (\mathsf{g}_o + \mathsf{g}_e . j) \rbrace,  \\
    E[i,1] &= F[1,j] = 0
    \end{split}
\end{equation}

Since BSW extends a seed, the initial matrix score $H[0,0]$ is the match score of the seed. Given the computation of a cell is only dependent on the three neighbors, BSW in BWA-MEM only maintains one row each of $E$, $F$, and $H$. Smaller sequence sizes ensure that these arrays are present in caches and only the sequence pairs need to be fetched from memory, making BSW a compute bound problem.

\subsection{Vectorization approaches and challenges}

Given that BSW is compute bound, we chose to vectorize it using SIMD parallelism. Vectorization of standard SW has been studied extensively. There are two popular approaches: intra-task  and inter-task. We see the following challenges in application of these approaches to BSW.

\subsubsection{Intra-task}

Intra-task approach vectorizes the computation of SW of one pair. It either leverages the independence between anti-diagonal cells of the matrix to vectorize across them or uses striped SIMD approach. However, this kind of approach is unwieldy for BSW. Short sequences, computation only within the band, and potential reduction in the band size due to band adjustment, restrict the available parallelism for vectorization.

\subsubsection{Inter-task}

In inter-task approach, each vector lane computes the matrix for different sequence pairs. The irregularity of matrix sizes, number of cells computed, and the positions within matrices of cells computed across matrices, makes extracting performance using inter-task approach difficult. 

Moreover, in BWA-MEM, whether a seed is extended using BSW or not depends on the previously extended seeds of a read. Such dependency restricts the parallelism across seeds of a read. 

\begin{figure}
    \centering
    \includegraphics[scale=0.3]{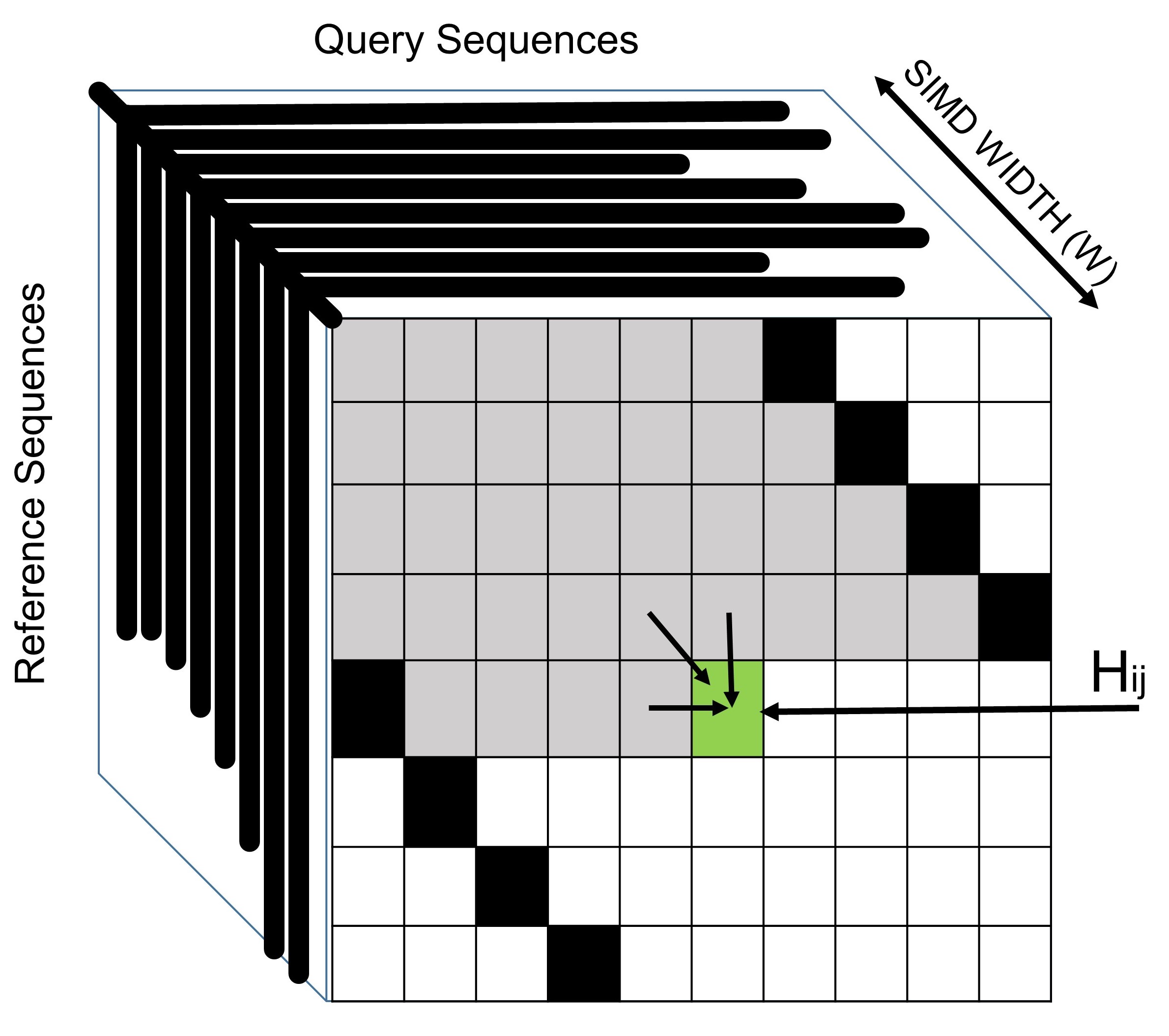}
    \caption{Inter-task vectorization of BSW over a batch of sequence pairs. Dark cells represent the banding, grey cells represent the computed cells in the matrix, and green cell represents the cell $H_{ij}$ being computed.}
    \label{fig:bsw}
\end{figure}

Due to availability of better parallelism, we utilize inter-task vectorization for accelerating BSW (Figure~\ref{fig:bsw}). 

\subsection{Inter-Task Vectorization}
\label{sec:opt-bsw-simd}

Our approach for inter-task vectorization is illustrated in Figure~\ref{fig:bsw} and is based on our previous work~\cite{misra2018}. Let $W$ be the SIMD width. We process $W$ sequence pairs at a time. For each cell $(i,j)$ that is computed, each vector lane computes the corresponding cell $(i,j)$ for all matrices. Therefore, even if one of the sequence pairs requires a particular matrix cell to be computed, the corresponding matrix cell is computed for all the sequence pairs leading to wasteful cell computations.

\subsubsection{Sorting}
While any aborted sequence pair can be replaced with a new one to avoid idle vector lanes, the overhead of the operation supersedes its benefits.
Wasteful cell computations due to variations in sequence lengths can be better curbed by ensuring uniformity in sequence lengths. We use radix sort to sort the tasks by their respective sequence lengths, and then group together tasks with the same or close sequence lengths to ensure uniformity of tasks filling vector lanes.

\subsubsection{Increasing parallelism}

Given the dependency between seeds, the parallelism is only available across reads. However, the count of seeds per read can vary a lot leading to imbalance of computation across reads. A dynamic allocation of reads to vector lanes can potentially balance computation across vector lanes. However, given the imbalance in number of computed seeds across reads, it requires a large number of reads in a batch to succeed. Large batches are not possible due to memory constraints as we have to maintain the metadata corresponding to each read between steps of the BWA-MEM algorithm. 
Therefore, we manage the dependency between the seeds by first extending all the seeds of a read, and then post process them to filter out the ones that should not have been extended.
 
\subsubsection{AoS to SoA conversion of sequence pairs}

We convert the input sequences from AoS to SoA format to allow reading of corresponding bases from $W$ sequence pairs using vector \textit{load} operation instead of a \textit{gather} operation.

\subsection{Overview of operations}

For a given row, we do the following computations. 
(a) To perform \textit{cell computations} within a given band, we utilize instructions such as \textit{cmp}, \textit{blend}, \textit{max}, \textit{mov}, \textit{add}, and \textit{sub}. Similar instructions are used to identify the best score and its position.
(b) To find the global score, at each cell in a row, we check whether the end of query is reached for all the pairs using \textit{cmp} instruction and update the global score and its position using \textit{cmp} and \textit{blend} instructions.
(c) After the row computations, the cell range (for next row) is adjusted again based on the number of empty cells from both ends of the row.
From both ends of the row, with one cell at a time, we use \textit{cmp} and \textit{blend} vector instructions to check for empty cells and update the range.
(d) Moreover, we use \textit{mask} and \textit{cmp} instructions for maintaining the correct values for aborted sequence pairs.

\subsubsection{Precision}

Smaller integer-widths provide more vector lanes.
Since matrix scores are proportional to sequence lengths, it is possible to use different integer-widths for matrix computation depending on sequence lengths. In BSW optimization, we use $8$-bit or $16$-bit implementations depending on the sequence length.


\section{Results}
\label{sec:results}

\blfootnote{Software and workloads used in performance tests may have been optimized for performance only on Intel microprocessors. Performance tests, such as SYSmark and MobileMark, are measured using specific computer systems, components, software, operations and functions. Any change to any of those factors may cause the results to vary. You should consult other information and performance tests to assist you in fully evaluating your contemplated purchases, including the performance of that product when combined with other products. For more information go to www.intel.com/benchmarks.

Benchmark results were obtained prior to implementation of recent software patches and firmware updates intended to address exploits referred to as "Spectre" and "Meltdown".  Implementation of these updates may make these results inapplicable to your device or system.}

\subsection{Experimental Setup}

\subsubsection{System Configuration}
\begin{table}[!t]
\caption{System configuration}
\label{table:sys-conf}
\centering
\begin{tabular}{lcc}
\hline
& \partCPUtm & \partCPUtm \\
& Platinum & E5-2699 \\
& 8180 Processor &   v3 Processor \\
& (SKX) &   (HSW) \\
\hline
Sockets $\times$ Cores $\times$ Threads& $2 \times 28 \times 2$ & $2 \times 18 \times 2$\\
AVX register width (bits) & $512$, $256$, $128$ & $256$, $128$\\
Vector Processing Units (VPU)& $2$/Core & $2$/Core \\
Base Clock Frequency (GHz) & $2.5$ & $2.3$\\
L1D/L2 Cache (KB) & $32/1024$ & $32/256$ \\
L3 Cache (MB) / Socket & $38.5$ & $45$\\
DRAM (GB) / Socket & $96$ & $64$\\
Bandwidth (GB/s) / Socket & $114$ & $68$\\
\hline\hline
Compiler Version & \multicolumn{1}{c}{ICC v. 17.0.2} & ICC v. 17.0.2 \\
\hline
\end{tabular}
\end{table}

While our implementation achieves best performance by running on latest processors with AVX512 support from heavy use of AVX512 instructions, we also support alternate AVX2-only and scalar-only executions for older generation multicore processors. To confirm the wide applicability of our improvements, we evaluated our implementations on two generations of processors -- the latest Intel Xeon Scalable family of processors (Skylake) and Intel E5 v3 family (Haswell). Our specific Skylake and Haswell processors are detailed in Table ~\ref{table:sys-conf} and referred to as SKX and HSW, respectively, from here on. Each AVX512 VPU can process multiple $8$-bit and $16$-bit integers thus capable of SIMD widths of $64$ and $32$, respectively; similarly, each AVX2 VPU can process multiple $8$-bit and $16$-bit integers thus capable of SIMD widths of $32$ and $16$, respectively. We used \Intel \VTunetm 2018 for performance analysis by measuring hardware performance counters.

We perform all our experiments on a single sockets of SKX ($28$ cores) and HSW ($18$ cores) and use \textit{numactl} utility to force all memory allocations to one socket. For multi-threaded runs, we run $2$ threads per core to get the benefit of hyper threads. Since optimizing file IO is beyond the scope of this paper, we do not include file IO time in any of our results.

\subsubsection{Datasets}
To demonstrate the performance of our improvements, we make use of the following real datasets. We use the first half of the human genome (version Hg38), containing nearly $1.5$ billion nucleotides, as the reference sequence for all our experiments. We use five different read datasets covering all prominent short read lengths (Table~\ref{tab:app_sn_dataset}) to demonstrate the versatility of our implementation. We received the read datasets D1 and D2 directly from the Broad Institute to evaluate the performance of our implementation.

\begin{table}[h]
    \centering
    \caption{Real \textit{read} datasets for application performance evaluation. 
    }
    \begin{tabular}{c|ccc}
    \toprule
        Dataset & \#Reads &  Read Length & Dataset Source \\
         \midrule
         D1 & $5 \times 10^5$ & 151 & Broad Institute\\
         D2 & $5 \times 10^5$ & 151 & Broad Institute\\
         D3 & $1.25 \times 10^6$ & 76  & NCBI SRA: SRX020470 \\
         D4 & $1.25 \times 10^6$ & 101 & NCBI SRA: SRX207170 \\
         D5 & $1.25 \times 10^6$ & 101 & NCBI SRA: SRX206890 \\

         \bottomrule
    \end{tabular}
    \label{tab:app_sn_dataset}
\end{table}

\subsubsection{Correctness}

For each experiment, we verified that our output is exactly identical to that of original BWA-MEM.

\subsection{Performance Evaluation of Key Kernels On a Single Thread}
In this section, we evaluate the benefits of our improvements on the individual key kernels using real datasets on SKX. We extracted the kernels from both the implementations so that we are comparing only the kernels and nothing else.

\subsubsection{SMEM search using FM-index}
\label{sec:smem-vtune}

\begin{table}[!t]
    \centering
    \caption{Evaluation of SMEM kernel on a single thread of SKX using performance counters. Read dataset: 60,000 reads from D2.}
    \begin{tabular}{l|rrr}
    \toprule
        Performance Counters & Original &  Optimized minus & Optimized \\
         &  &  S/W prefetching &  \\
         \midrule
        \# Instructions ($\times$ $10^6$)	& 17,117	& 7,880	& 8,160 \\
		\# Loads ($\times$ $10^6$)	& 4,429	& 2,200	& 2,115 \\
		\# Stores ($\times$ $10^6$)	& 1,696	& 1,415	& 1,393 \\
		\# LLC Misses ($\times$ $10^6$)	& 23.9	& 29.7	& 9.5 \\
		Average latency (cycles) & 24 & 33 & 18 \\
		\# Cycles	consumed ($\times$ $10^6$) & 10,496	& 6,986	& 5,238 \\
        \midrule
		Time &	4.20s	& 2.79s	& 2.10s \\
        \bottomrule
    \end{tabular}
    \label{tab:smem-counters}
\end{table}


Table ~\ref{tab:smem-counters} evaluates our improvements to the SMEM kernel. Our modifications in the design to use the value of bucket size $\eta$ as $32$ and vectorization to reduce the number of instructions has resulted in a greater than $2\times$ reduction in the total number of instructions and the number of load instructions. 
However, lower value of $\eta$ results in less values of $k$ and $k+s$ falling into the same bucket, thus resulting in lower cache reuse. This is shown in the increase in the number of LLC misses and average latency of memory accesses between ``Original'' and ``Optimized minus S/W prefetching'' implementations. The use of software prefetching dramatically reduces the number of LLC misses (by nearly $3\times$) and average latency of memory accesses at the cost of slightly higher instruction count. The change in performance counters is clearly reflected in the time consumed. Overall, our improvements achieved $2\times$ performance gain over the original.

\subsubsection{Suffix Array Lookup (SAL)}
\label{sec:sal-vtune}

\begin{table}[!t]
    \centering
    \caption{Evaluation of SAL kernel on a single thread of SKX using performance counters. We prepared the input data for this by running the full application using real data and extracting the input to this stage. Read dataset: 600,000 reads from D2.}
    \begin{tabular}{l|rr}
    \toprule
        Performance Counters & Original &  Optimized\\
         \midrule
        \# SA offsets ($\times$ $10^6$) & \multicolumn{2}{c}{41.05} \\
        \midrule
        \# Instructions ($\times$ $10^6$)	& 213,065	& 1,058 \\
		\# Loads ($\times$ $10^6$)	& 23,617	& 241 \\
		\# Stores ($\times$ $10^6$)	& 5,459	& 156 \\
		\# Inst. per SA offset &	5,190.7	& 25.8 \\
		\# LLC Misses ($\times$ $10^6$)	& 452.3	& 5.0 \\
		Average latency (cycles) & 26 & 69 \\
		\# Cycles	consumed ($\times$ $10^6$) & 161,170	& 880 \\
        \midrule
		Time &	64.47s	& 0.35s \\
        \bottomrule
    \end{tabular}
    \label{tab:sa2ref-counters}
\end{table}


Table ~\ref{tab:sa2ref-counters} shows the benefit of simplifying the SAL kernel. There is a dramatic reduction in the number of instructions per SA offset (nearly $200\times$). Therefore, despite the increase in number of LLC misses per load, and thereby average memory latency, we see a speedup of nearly $183\times$. There is further scope of performance improvement by using software pre-fetching. However, for our optimized implementation the time consumed in this kernel is negligible compared to the full application run time. Therefore, there isn't much benefit in pursuing it.

\subsubsection{Banded Smith-Waterman}
\label{sec:bsw-vtune}


\begin{table}[!t]
    \centering
    \caption{Run-time of optimized (with AVX512) and original BSW benchmarks on a single thread of SKX for $48$ million sequence pairs obtained by running the full application using real data and intercepting the input to BSW stage. Read dataset used: D3.}
    \begin{tabular}{c|c|cc|cc}
        \toprule
         \multirow{2}{*}{BSW Benchmarks} & Original & \multicolumn{2}{c|}{16-bit} & \multicolumn{2}{c}{8-bit} \\
         & scalar & w/o sort & w/ sort & w/o sort & w/ sort\\
         \midrule
         Time (sec) & 283 & 65.36 & 44.46 & 42.09 & 24.46 \\
         \bottomrule
    \end{tabular}
    \label{tab:bsw_ts}
\end{table}

Table \ref{tab:bsw_ts} shows the performance benefit of our vectorized implementations at $8$-bit and $16$-bit precisions with or without sorting. For this experiment, we only used the sequence pairs for which $8$-bit precision was sufficient. The results show clear benefit of sorting which provides $1.5\times$ to $1.7\times$  performance boost by increasing the chances of pairs that are processed simultaneously using SIMD parallelism having similar lengths. Overall, our $8$-bit and $16$-bit implementations achieve nearly $11.6\times$ and $6.7\times$ speedup, respectively, over the original implementation.

\begin{table}[!t]
    \caption{Evaluation of optimized 8-bit (with AVX512) and original  implementations of BSW benchmark on a single thread of SKX using hardware performance counters.}
    \centering
    \begin{tabular}{l|cc}
    \toprule
         Performance Counters & Original & Optimized 8-bit \\   
        \midrule
         \# Instructions &  $1,385 \times 10^9$ & $100 \times 10^9$\\
         \# Clock cycles &  $440 \times 10^9$ &     $46 \times 10^9$\\
         IPC & $3.14$ & $2.17$ \\
         \bottomrule
    \end{tabular}
    \label{tab:vtune}
\end{table}

Table ~\ref{tab:vtune} uses hardware performance counters to explain the speedup. BSW is a compute-intensive task.
Consequently, any reduction in number of instructions executed would result in reduction in execution time. 
Vectorization led to reduction in the number of instructions executed by $13.85\times$ (for $8$-bit implementation) over the original implementation of the BSW kernel that was a scalar implementation. The reduction in IPC is explained by the fact that a large majority of the instructions in our optimized code are SIMD instructions. Therefore, with two ports for SIMD instructions, an IPC of $2.17$ is expected while accounting for a few scalar instructions. In contrast, there are $4$ ports for scalar ALU instructions explaining an IPC of $3.17$ for the original implementation. 

\begin{table}[!t]
    \centering
    \caption{Breakup to run time of optimized 8-bit BSW (with AVX512) on a single thread of SKX.}
    \begin{tabular}{l|r}
    \toprule
        Components & Time (\%)  \\
        \midrule
        Pre-processing    & 33 \\
        Band adjustment I & 9 \\
        Cell computations & 43 \\
        Band adjustment II & 15 \\
        \bottomrule
    \end{tabular}
    \label{tab:bsw_perf_anal}
\end{table}

Note that $16$-bit and $8$-bit integer precision values provide SIMD widths of $32$ and $64$, respectively, with AVX512 facilitating potential speedups of $32\times$ and $64\times$, respectively. In order to understand the gap between these ideal speedups and achieved speedups, Table ~\ref{tab:bsw_perf_anal} presents the breakdown of time spent in different components of the optimized $8$-bit BSW. 
Run-time profiling shows that only $43\%$ of the total time is spent in calculating the DP matrix; while remaining time is spent in pre-processing of the input sequences (which converts the input sequences from AoS to SoA order), and cell range adjustment. AoS to SoA conversion is performed specifically for the vectorized implementation and is not needed for the original scalar implementation.

Owing to multiple factors such as variations in sequence lengths and aborting of matrix computations (discussed in section~\ref{sec:opt-bsw}), our inter-task vectorization ends up computing wasteful cells.
Empirical analysis shows that the useful cells are roughly half of the total cells computed. Therefore, only $21.5\%$ of the run time is spent in computation of useful cells of the DP matrices.


\subsection{End-to-end Performance Evaluation on Single Socket}


In this section, we present the comparison of the end-to-end compute time of the original and our optimized implementation of BWA-MEM, with all the optimized kernels integrated.

\subsubsection{Scaling with respect to the number of cores}

\begin{figure}[!t]
    \centering
    \includegraphics[width=\columnwidth, height=4cm]{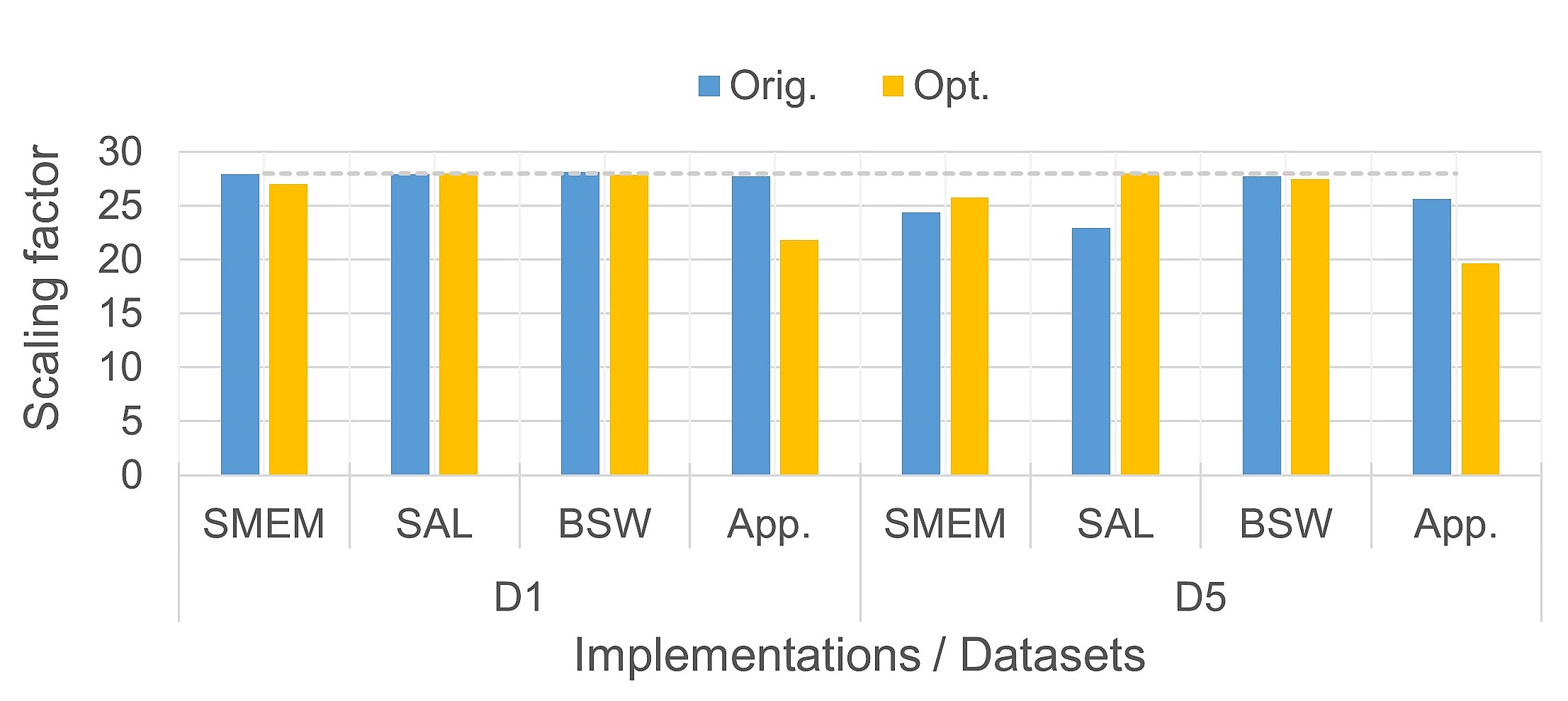}
    \caption{Comparison of scaling of three kernels and application between our implementation (Opt.) and original BWA-MEM (Orig.), from 1 core to 28 cores of SKX. Dotted line represents the perfect scaling. Datasets used: D1 and D5. Turbo boost was switched OFF for reliability of scaling performance.} 
    \label{fig:sn_app_scal}
\end{figure}

We evaluate the multicore scaling of our implementation and three integrated kernels in (Figure~\ref{fig:sn_app_scal}). All the kernels demonstrate good scaling from 1 core to 28 cores on single socket closely matching the scaling of the original BWA-MEM with our optimized implementation achieving better scaling for the memory intensive SMEM and SAL kernels for the D5 dataset. We see a dip in scaling performance of the overall application for our implementation (D1: $22\times$, D5: $20\times$). The decline in scaling stems from the part of the application other than the three kernels. That is the part of the implementation that has not been optimized yet. Though all three key kernels achieve greater than $25\times$ scaling, some of the components in \emph{Misc} are memory bandwidth bound and do not scale beyond $15\times$.

\subsubsection{Time to solution}

\begin{figure*}[!t]
\begin{minipage}{\linewidth}
{
	\subfloat[Single thread of SKX]
	{
		\includegraphics[width=0.5\linewidth]{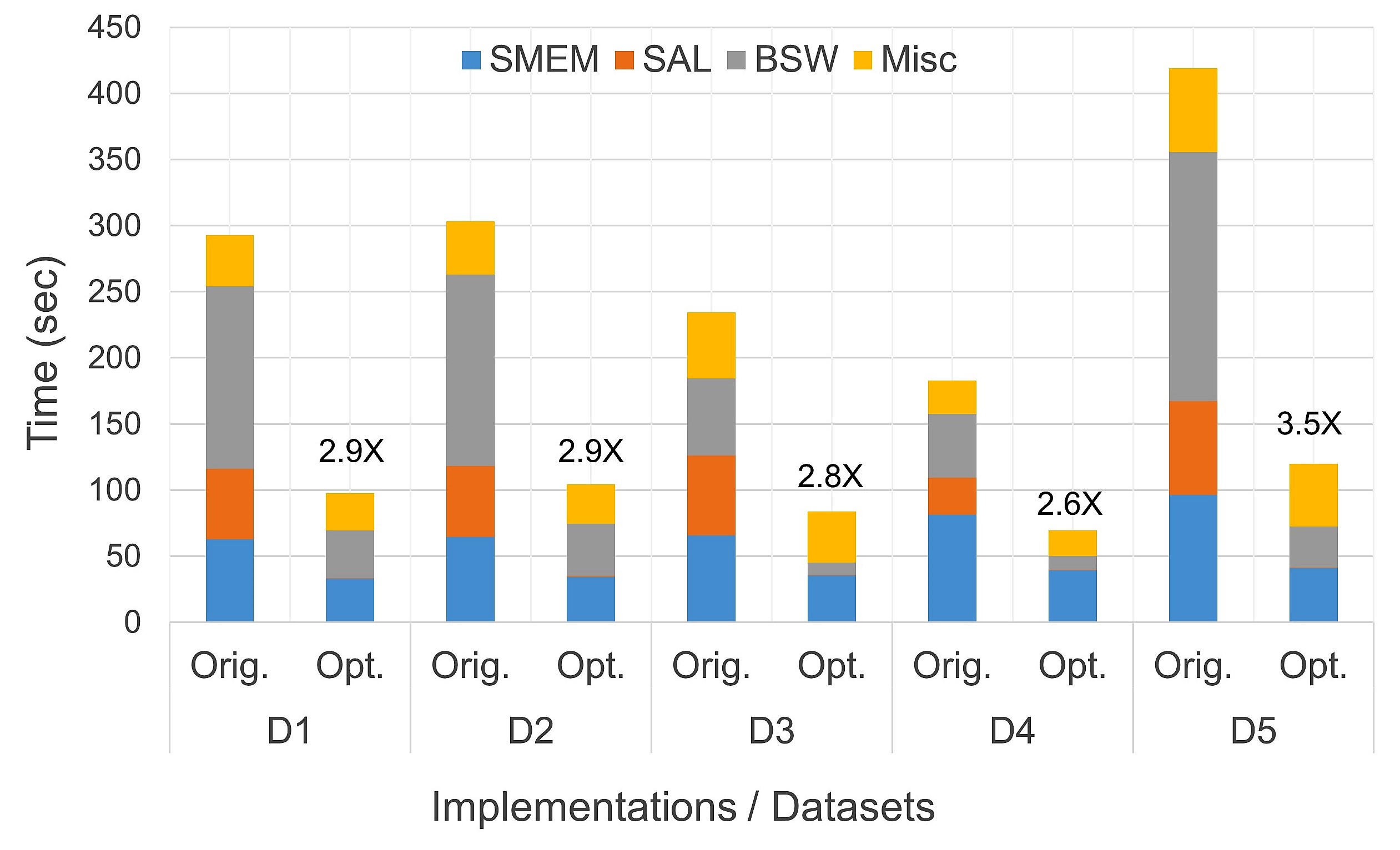}
		\label{fig:sn_st1}
	}
	\hfil\hfil
	\subfloat[Single socket (56 threads/28 cores) of SKX]
	{
		\includegraphics[width=0.5\linewidth]{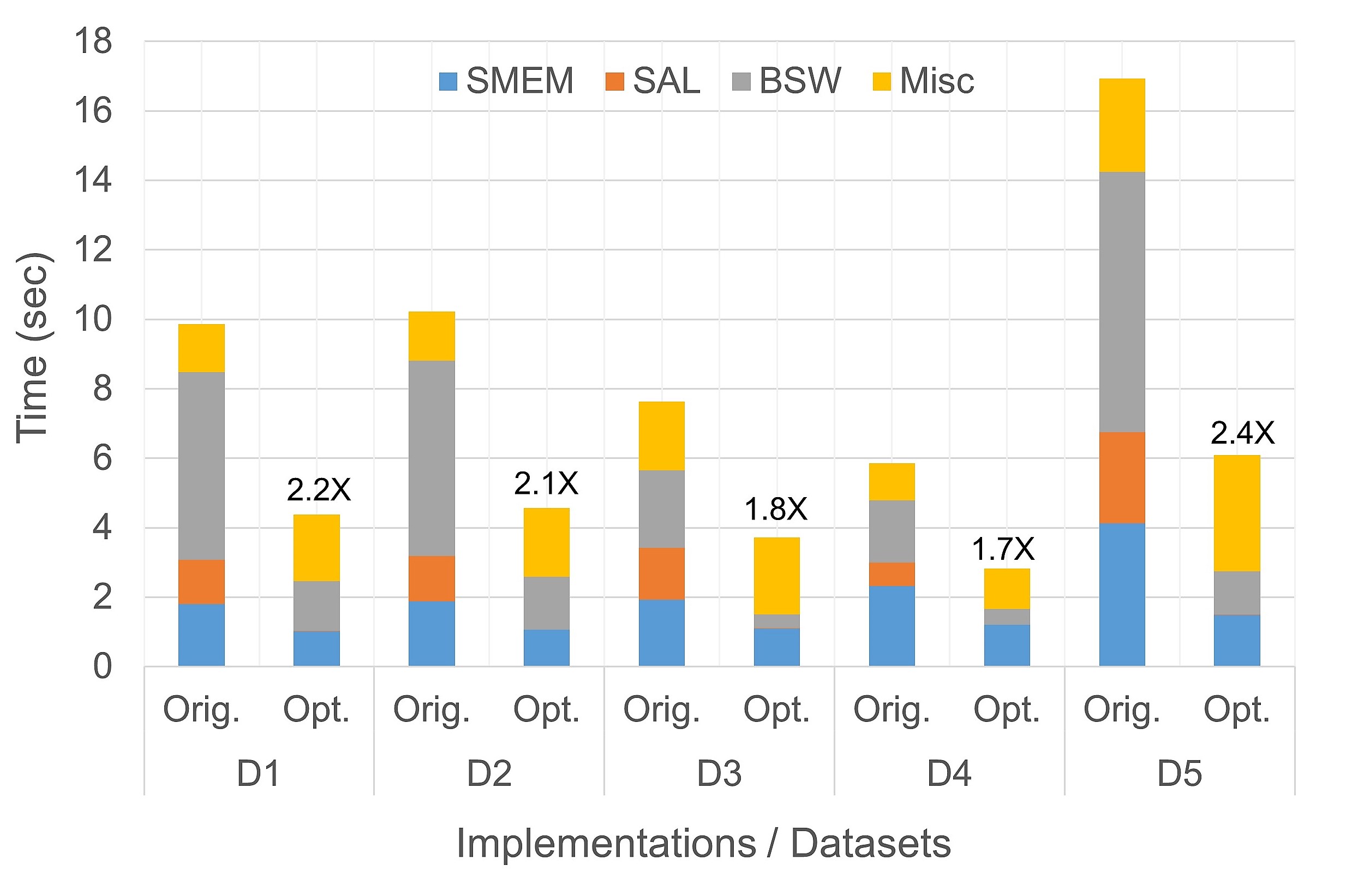}
		\label{fig:sn_ts56}
	}
	\vfill
	\subfloat[Single thread of HSW]
	{
		\includegraphics[width=0.5\linewidth]{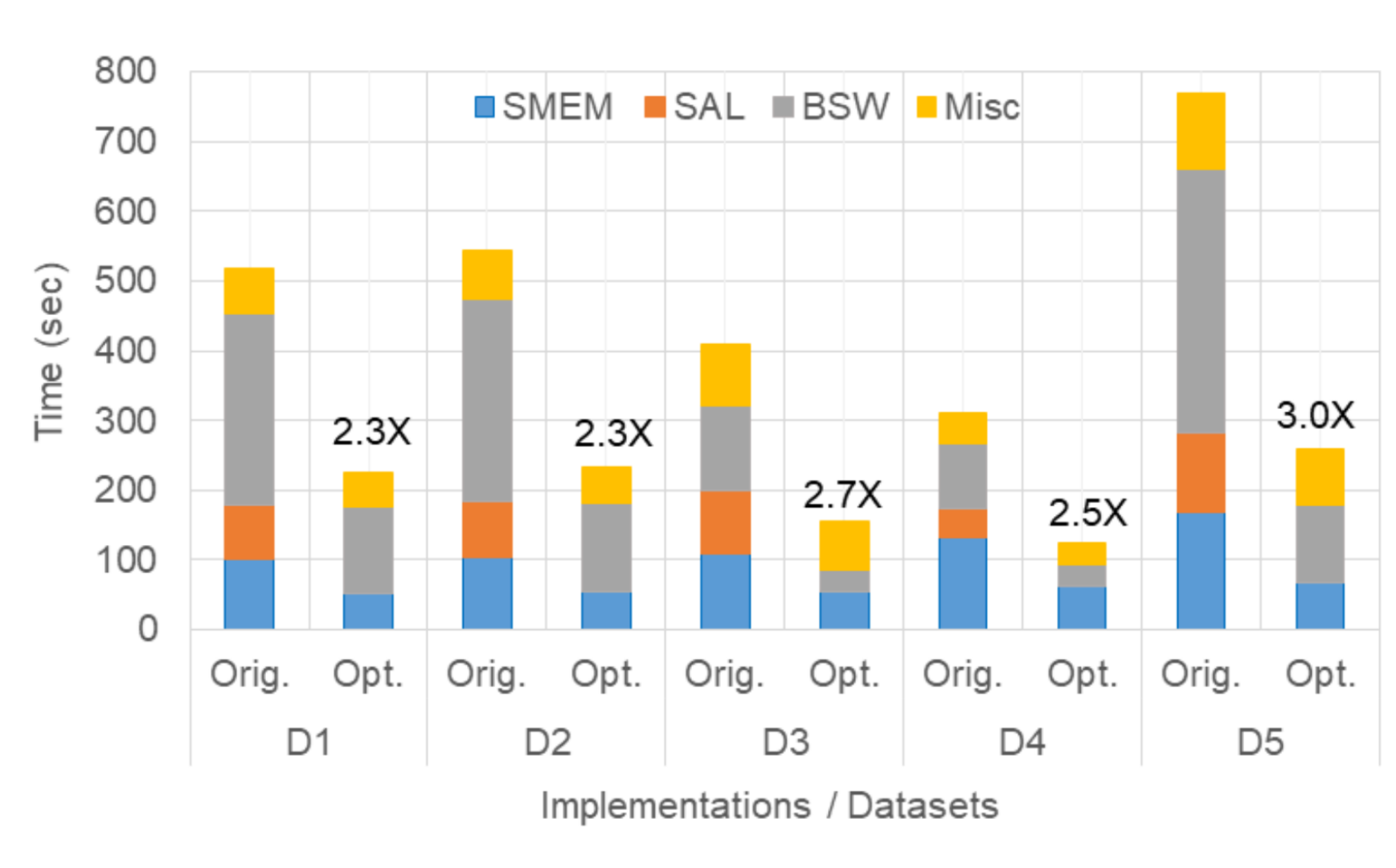}
        \label{fig:sn_st1_hsw}
	}
	\hfil\hfil
	\subfloat[Single socket (36 threads/18 cores) of HSW]
	{
        \includegraphics[width=0.5\linewidth]{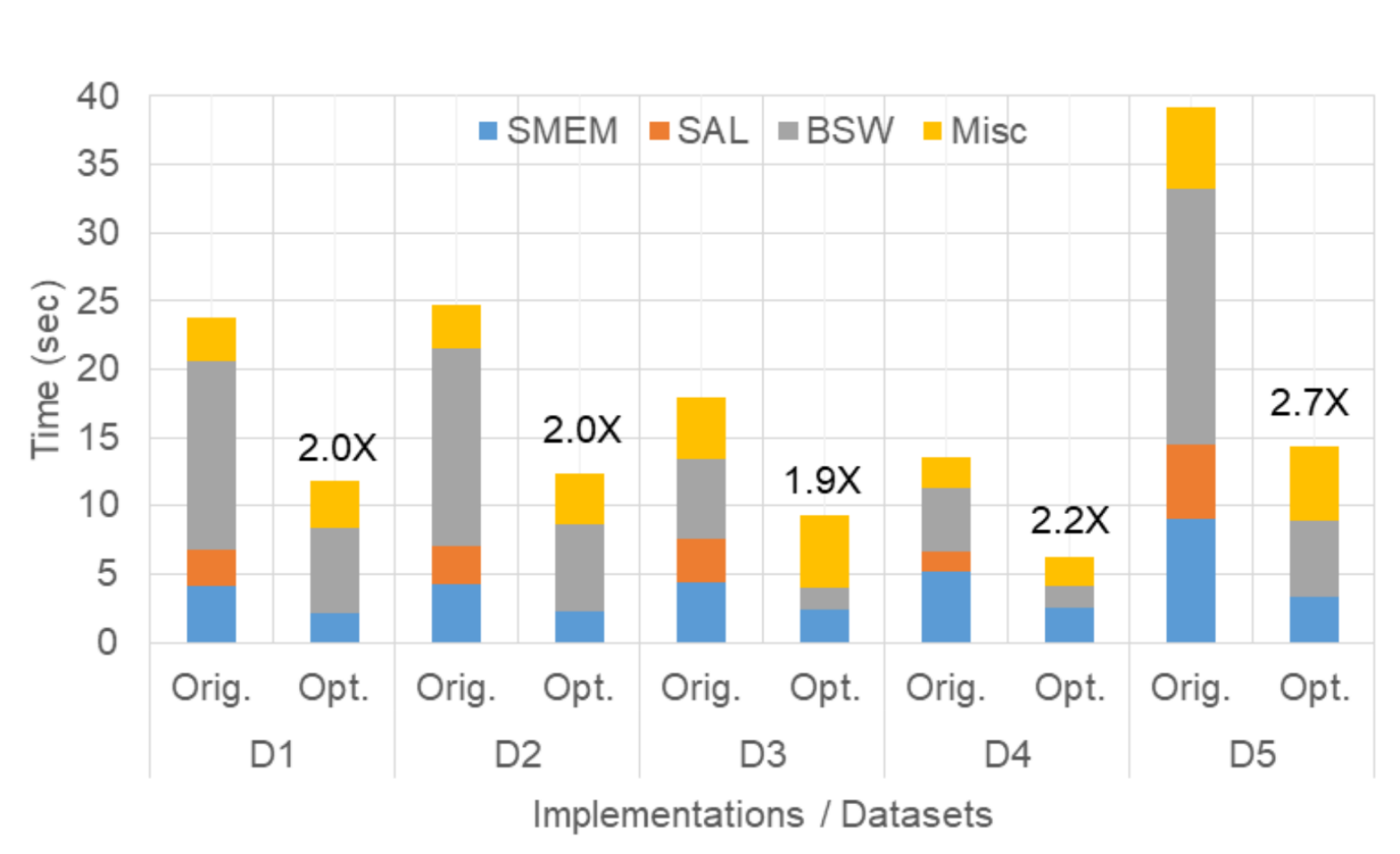}
        \label{fig:sn_ts36_hsw}
	}
}
\end{minipage}
\caption{Comparison of performance of BWA-MEM (Orig.) and our solution (Opt.) on single thread and single socket of SKX and HSW on five real datasets. Misc is the run-time consumed by non-kernel part of the implementation. Performance numbers at the top of Opt. bars represent speedups of our approach over BWA-MEM.}
\label{fig:sn}
\end{figure*}

We compared the time to solution of our optimized implementation with the original implementation using five different real datasets (Table~\ref{tab:app_sn_dataset}). 
The run-time of our two implementations and original BWA-MEM, and speedup of our implementations over original BWA-MEM, are depicted in Figures~\ref{fig:sn_st1} \&~\ref{fig:sn_st1_hsw} (single thread) and Figures~\ref{fig:sn_ts56} \&~\ref{fig:sn_ts36_hsw} (single socket).

All the three kernels sustain their benchmark speedup in the application. In fact, SAL kernel's run-time barely contributes to the overall run time due to the $183\times$ speedup. However, application performance benefit for the BSW kernel is lowered owing to the processing of extra seeds (Section ~\ref{sec:opt-bsw-simd}). On an average, we see $14\%$ extra sequence pairs aligned by our method. 
For dataset D2 ,on SKX, we observed that there were $13.5\%$ extra seed pairs, but resulted in nearly $1.43\times$ more BSW time, incurring a loss of speedup over BSW of BWA-MEM by $1.47 \times$.
On overall application: on SKX, we see speedup range of $2.6\times - 3.5\times$ on single thread, and of $1.7\times - 2.4\times$ on single socket; while, we see speedup range of $2.3\times - 3.0\times$ on single thread, and of $1.9\times - 2.7\times$ on single socket, on HSW.






%


\section{Related Work}

Owing to the complexity of BWA-MEM, a majority of the attempts to speed it up are confined to accelerating one of the three kernels, either on GPGPU or on FPGA~\cite{houtgast2018, chang2016, ahmed2015}. 
Additionally, these methods pipeline the non-optimized kernels on the host CPU for execution. Out of the three kernels, given that standard SW has been extensively targeted for acceleration and BSW of BWA-MEM is a variant of standard SW, a lot of studies~\cite{houtgast2018, ahmed2015} have targeted only BSW for acceleration. Due to dependencies between the seeds within a read, all of the mentioned acceleration approaches for BSW kernel employ intra-task parallelism.
Excluding ~\cite{ahmed2015}, the other approaches manage to achieve $1.6\times$ to $3\times$ speedup on BSW and $1.45\times$ to $2\times$ overall. 
Chang~\textit{et al.}~\cite{chang2016} exclusively target SMEM  optimizations on FPGAs and demonstrate $4\times$ and $1.26\times$ performance improvement on SMEM and overall application, respectively. Due to  pipelining of computations between host CPU and accelerator, the benefits of BSW speedup are contingent on speed of non-optimized kernels. 
Ahmed \textit{et al.}~\cite{ahmed2015} target all the three kernels for optimizations. They use $4$ FPGAs to speedup SAL and BSW kernels by $2.8\times$ and $5.7\times$ respectively. Also, they apply algorithmic enhancements to optimize SMEM kernel on host CPU, speeding it up by $1.7\times$ and overall performance by $2.6\times$.
However, their output may differ from BWA-MEM output, due to bypassing of some of the heuristics used in BWA-MEM BSW. To the best of our knowledge, no published work contains a holistic architecture-aware optimization of BWA-MEM software on multicore systems.


FM-index methods and SW based alignment are widely used techniques in NGS data analysis and have been classified as computational building blocks~\cite{vasim-bb-bioaxiv-2018}.
Due to the intricacy of FM-index based algorithms, few works considered optimization of these problems \cite{zhang2013optimizing,bwt-chacon-2015,nstep-chacon-2013, nvbio,rastogi2014, nunes2015} and most of them have targeted the simpler, exact search algorithm (find exact matches of full query sequences).
Chacon {\it et al.}~\cite{nstep-chacon-2013} developed a clever technique for exact search, which reduces the number of memory accesses. Application of their technique to SMEM search is non-trivial. Additionally, they tried software prefetching for exact search~\cite{nstep-chacon-2013} with minimal benefit due to an inefficient design. 
Misra {\it et al.}~\cite{misra2018} demonstrated significant performance gains on exact search with software prefetching by batching multiple queries together to hide memory latency.

Intra- and inter-task vectorization are both popular approaches for SW acceleration~\cite{misra2018, Daily2016, rognes2011, Li2007, liu2010}. However, BSW of BWA-MEM has some key variations from standard SW as discussed in section~\ref{sec:opt-bsw}. Therefore, there have been only a few attempts to accelerate exact BSW kernel from BWA-MEM.

\section{Conclusion and Future Work}
\label{sec:conclusion}

In this paper, we presented an efficient implementation of BWA-MEM for multicore systems while maintaining identical output so that the current users of BWA-MEM can seamlessly switch to the new implementation. We presented our improvements through architecture aware optimization to speedup the three key kernels - SMEM, SAL and BSW - by $2\times$, $183\times$, and $8\times$, respectively. This results in an overall performance gain of $3.5\times$ and $2.4\times$ over the compute time of BWA-MEM on a single thread and single socket, respectively, of an Intel Skylake processor. 
Our implementation has been open sourced so that the users of BWA-MEM can benefit from increased performance.

We were successful in making the SAL kernel virtually insignificant contributor to the overall run time. The other two kernels are instruction bound and SMEM is also partly memory latency bound. Given the irregular structure of the algorithms, SMEM can not benefit from vectorization, while gains for BSW due to vectorization are limited. Therefore, processors that rely heavily on SIMD performance can only achieve limited performance on BWA-MEM. While better support for gather operations will help, an architecture that does not rely on SIMD for performance will be better.

As future work, we plan to make the following improvements - (i) 
attempt to improve the parts of BWA-MEM apart from the three kernels that we focused on in this paper in order to improve multicore scaling and performance of the overall application further, and (iii) explore if there is additional scope of reducing average memory latency of SMEM kernel and the instruction counts of SMEM and BSW kernels.

\bibliographystyle{IEEEtran}
\bibliography{bwa-mem2}

\end{document}